\documentclass[amssymb,twocolumn,aps,showpacs]{revtex4}

 \usepackage{epsfig}
\usepackage{dcolumn}

\begin{document}

\title{Symmetry Analysis of the Kohn-Sham Band Structure of Bulk Lithium Fluoride}

\author{Richard J. Mathar}
\email{mathar@mpia.de}
\homepage{http://www.mpia.de/MIDI/People/mathar}
\affiliation{
Goethestr.\ 22, 69151 Neckargem\"und, Germany}

\date{\today}

\begin{abstract}
Kohn-Sham orbitals of face-centered cubic lithium fluoride
are calculated in prototypical local-density approximations
to the exchange-correlation functional. The symmetry
analysis of these Bloch functions in a LCAO basis
on a path $\Gamma$-X-W-K-$\Gamma$-L-W through the Brillouin
Zone is compiled into a list of errata to symmetry labels
in the LiF literature, the bulk of which dates back
to the 1970's.

\end{abstract}
\pacs{61.50.Ah, 71.20.Ps, 71.15.Mb}
\maketitle

\section{Scope}
The crystallographic symmetry of the ground state of solid lithium fluoride (LiF) is
simple: a face-centered cubic Bravais lattice with one formula unit in the
primitive unit cell, known as the rock salt structure, space group $Fm3m$ and
number 225 in the International Tables \cite{Hahn}.
Yet the symmetry assignments to the lines and points of symmetry of
(effective) single-particle orbitals (wavefunctions) populating
the reciprocal unit cell vary in the literature to a high degree, with no
discussion that would unify the diverging opinions.
This work is to be considered an extended erratum to
consolidate the graphs of band structures of LiF published in the past
that are incoherent or violate rules of
the compatibility tables.

We start from self-consistent Kohn-Sham (KS) orbitals obtained
in a Local Density Approximation (LDA) to the exchange-correlation (XC) functional
of the Density Functional Theory (DFT),
and apply the space group operators to points of low and high symmetry
in the Brillouin Zone (BZ), which lead to symmetry labels read
from the character tables of their little groups. Band structure line
graphs result as we connect the Kohn-Sham eigenvalues
in accordance with the compatibility tables. 

\section{Kohn-Sham orbitals in the LDA}
\subsection{GTOFF Bloch Functions}

The basis of this work are the LiF eigenvectors (Bloch functions) and eigenvalues
(energies) calculated by the non-relativistic version of GTOFF \cite{BoettgerPRB53,BoettgerIJQC60,BoettgerIJQCS29},
which solves the self-consistent Kohn-Sham equations with an all-electron LCAO ansatz
for the Bloch functions, and builds the atomic orbitals from linear combinations
of (Hermite) Gaussian Type primitives. We use its ``bulk'' option (translational
periodicity of the Kohn-Sham Hamiltonian in three directions) and two variants
of the LDA, one in the next section merely to demonstrate compatibility of the outcome
with other, independent calculations, and an older one in the main section
to make close contact to the Zunger--Fremann publication \cite{Zunger}, which, by
the number of citations received, could be called the ``reference.''

Unless noted otherwise, the results are calculated using a Kohn-Sham basis with
23 Gaussian Type Orbitals GTO's at the Li and 46 GTO's at the F sites (Table \ref{gtos}).

\subsection{LDA band gap}

The direct band gap $E_g$ at $\Gamma$
in the Hedin-Lundqvist-type Local Density Approximation (LDA) to the
exchange-correlation \cite{Moruzzi} is found to be
\[
E_g/\mbox{eV}\approx 8.99-3.41\Delta+1.24\Delta^2 \qquad
(\Delta\equiv a/a_0-7.5939)
\]
as a function of the lattice parameter $7.37a_0<a<8.05a_0$, where
$a$ is the edge length of the underlying simple cubic unit cell, where
$a/2$ is the shortest distance between Li and F, and $a_0$ the Bohr radius.
(The theoretical equilibrium lattice constant is smaller than $7.37a_0$,
hence underestimates the experimental 
value, $7.60841a_0$ at 25$^\circ$C \cite{Landolt}, by at least 3\%.
In a recent pseudopotential calculation \cite{Wang}, the deviation is 2.7\%.)
The value is compatible with the 8.7 eV of a pseudo-potential
calculation with the Ceperley-Alder LDA to the
exchange-correlation \cite{Shirley96} at $a=7.62a_0$, and values of 8.89 eV
(8.82 eV)
by an independent Full-Potential Linearized Augmented Plane-Wave
program at $a=7.59a_0$ ($7.61a_0$) for the same energy functional \cite{Blaha}.

The band gap of the X$_\alpha$ ($\alpha=2/3$) calculation is about $1.0$ eV
smaller
than calculated by Zunger and Freeman \cite{Zunger}, which probably 
indicates that their method of calculation starts from parameters of
isolated ions and does not achieve full self-consistency at the end.
(Their program likewise seemed to over-estimate the direct band gap of cubic
boron nitride \cite{ZungercBN} by $\approx 2$ eV compared with the consolidated
theoretical result \cite{Kim,Hernan,Furthm,Park} of 8.8 eV, and of diamond \cite{Zungerdia}
by $\approx 0.7$ eV compared with other theoretical results
\cite{Salehpour,Surh,Willatzen,MatharDia}.)

A considerably smaller theoretical LiF band gap of 7.65 eV has been reported by
Ching {\it el al.\/} \cite{Ching} for $a=7.591a_0$.
Assessment of this number is not clear; the optical band gap of MgO is
given as 5.13 eV \cite{Ching}, whereas an earlier value by what looks
like the same methodology is 4.19 eV \cite{Xu}.

\section{Symmetry Analysis}
\subsection{Character tabulation}
Point group labels are determined following \cite{Bradley},
including the Errata by Stokes and Hatch \cite{Stokes}.
(In addition, $\sigma_z$ ought be added for $X$ in
the $\Gamma_\mathrm{c}^\mathrm{f}$ lattice in \cite[Tab.\ 3.6]{Bradley}.
Evarestov and Smirnov interchange $B_1$ and $B_2$ for $C_{6v}$ in
\cite[Tab.\ 3.6]{Evarestov} relative to \cite[Tab.\ 2.2]{Bradley}.
This is not simply a typo but persists if we compare the $C_{6v}$
entry in \cite[Tab.\ 4.2]{Evarestov} with the $D_{6h}$ table
of Flurry \cite[App.\ 4]{Flurry}. This does not cause confusion here,
as $C_{6v}$ does not enter the subsequent analysis of a cubic space group.)

The computational procedure is to construct the character tables
of the ten abstract groups of the lines/points of symmetry in the Brillouin
Zone according to \cite[p.\ 389]{Bradley}.
An eleventh entry of the form $$O\quad {\bf G}_2^1:(\sigma_z,0)\ldots:c$$
is added to account for points in the $O$-plane \cite[Tab.\ 3.11]{Bradley}
that we meet walking from W to K\@.
GTOFF orbitals are rotated to ensure
their ${\bf k}$-points fall into the ``irreducible'' wedge of the
BZ defined in \cite[Fig.\ 3.14]{Bradley}.
For each class of the abstract group of the ${\bf k}$-point, a character
of the GTOFF orbital (Bloch function) $\langle {\bf r}|\nu {\bf k}\rangle$
of band $\nu$ is computed as the integral
$\sum_d \langle \nu {\bf k}|\{R|{\bf v}\}|\nu {\bf k}\rangle$,
applying one (any) Seitz operator $\{R|{\bf v}\}$ of this class.
The sum $d$ over up to three degenerate levels deals automatically with
some arbitrariness of the pointing directions of the eigenstates/eigenvectors
in these cases that has been left over from use of a (generic) eigenvalue routine
in the GTOFF solver.
Since GTOFF employs (Hermite) Cartesian Gaussians in the molecular basis,
the point operators $R$ are adequately represented by Cartesian,
matrices: Operation of these on the bases is a
multinomial re-expansion in the $x$, $y$, and $z$ coordinates \cite[Sec.\ 1.4]{Bradley}.
All operators $R$ are
members of the little co-group; therefore $\{R|{\bf v}\}|\nu{\bf k}\rangle$
is another Bloch function at the same ${\bf k}$, and the integrals (matrix
elements) become standard LCAO overlap integrals.

This determination of characters associated with 
a list of Seitz operators is numerically stable. Matrix representations
of the abstract groups are not needed. It does not work blindly
in cases of \cite[Tab. 5.7]{Bradley} where the authors chose generators of
the abstract group that are not space group operators: consider the example
of $Q^x$ of space group 227:
the translation ${\bf v}=(\frac{1}{4},\frac{1}{4},\frac{1}{4})$ has been
removed from $R=C_{2f}$, and the overlap integral cannot ``recover'' the
missing overall phase factor associated with this virtual relocation of the
rotated image. (In fact, all pair-wise overlap integrals become wrong caused
by incorrect distances between the Wyckoff positions in the ``bra'' and the ``ket.'')
The alternative is comparison of the Bloch function and its rotated image in
selected regions of the BZ, but this calls for more advanced methods of
``important sampling'' to bypass nodes of the orbitals or dispose regions of small orbital
density (in case of strongly localized core orbitals).

The representation of the abstract group that matches this list of characters
is translated to a point group symbol by \cite[Tab.\ 5.8]{Bradley}.
Subduction frequencies (compatibility tables) are generated from
the point group genealogy \cite[Fig.\ 4.1]{Bradley} and \cite[(4.3)]{Koster}
for adjacent points on the BZ path.

\subsection{Symmetry assignments and level crossing}
The band structure of Fig.\ \ref{figure1.ps} ensues in compliance with
the compatibility tables \cite{Cracknell,Bouckaert,Flurry,Salthouse,Koster}.
Lines are the linear (tetrahedral) interpolations/connections
obtained from the symmetry analysis; so there is no finesse in regions
where perturbation theory proves that there is no
linear term in the series expansion of the energy as a function of ${\bf k}$.
Point groups labels follow \cite{Bradley} and may be converted to
the convention of Bouckaert {\it et al.\/} \cite{Bouckaert}
using Tab.\ \ref{pglabels}.
The symmetry labels refer to a setting with a Li atom at
the origin of coordinates, and are not always unique.
The state near $-11$ eV at L, for example, consists
of $p$-orbitals on the F site plus some $s$-contribution on the Li site.
This Bloch-type KS orbital is even under inversion at {\em this\/}
origin.
If a F atom is chosen to define the origin of coordinates,
this same KS orbital is odd under inversion at this new origin, and
labels have to be swapped as follows \cite[Tab.\ 4.7]{Evarestov}\cite{Antoci}\cite[Fig.\ 5]{Ewing}:
A$_1\leftrightarrow$B$_2$, A$_2\leftrightarrow$B$_1$ (at W);
A$_{2u}\leftrightarrow$A$_{1g}$, E$_g\leftrightarrow$E$_u$ (at L);
A$\leftrightarrow$ B (at Q).

The solid lines in Fig.\ \ref{figure1.ps} are derived following a no-crossing rule
for energy dispersions of the same symmetry \cite{Herring}.
The rule applies perturbation theory to deduce that in almost all cases
there is some residual interaction build into the Hamiltonian that splits
energy levels that close up. This argument of \cite[Sec.\ 1]{Herring}
does not apply to the
cases of empty orbitals--the simplest demonstration are the level crossings
in empty lattices (Fig.\ 4.4.3 in \cite{Callaway})---that are eigenstates
of the full/complete (self-consistent) 1-particle Hamiltonian.
Our ground state Hamiltonian does {\em not\/} include any mechanism
(term) to lift two electrons to the crossing regions to modify the
energetically degenerate orbitals.
Dashed lines above 25 eV indicate some (but not all)
alternative choices, if ``accidental'' degeneracies
are not ruled out.
To decide which of these connections are ``correct,'' various approaches are helpful.
(i) According
to ${\bf k\cdot p}$ perturbation theory, the overlap
$\langle \nu {\bf k}|\nu ' {\bf k}+{\bf q}\rangle$ will be close to 1
in a neighbourhood of small ${\bf q}$ around a ${\bf k}$-value, if
the correct band $\nu=\nu '$ is chosen. In practice, however, the
density of the mesh in the reciprocal lattice hardly is dense enough 
to have sufficiently small $|{\bf q}|$ at hand.
(ii) One might compute the eigenvalues at an extra intermediate
${\bf k}$-point close to the expected crossing.
(iii) The connectivity
of bands does not change when the lattice is set up at a large lattice
constant (with almost horizontal bands and therefore a minimum of accidental
crossings), then is ``adiabatically'' contracted to its actual value. 
With this idea in mind we calculate the same crystal at a larger
lattice constant grown to 110\% (Fig.\ \ref{figure2.ps}). The prospective accidental
crossing at Q near 28 eV in Fig.\ \ref{figure1.ps} is not supported
by Fig.\ \ref{figure2.ps}, because the state A$_{2u}$ at L is still
connected to B$_2$ at W near 23 eV in Fig.\ \ref{figure2.ps}.
(This result is confirmed by the first approach, the inspection
and comparison of the dominant basis function of the four states.)
In contrast, 
the two accidental crossings near 32 eV at $\Lambda$ in
Fig.\ \ref{figure1.ps} seem actually to exist, as a
band of A$_1$ symmetry with very strong dispersion rapidly moves
upwards as $a$ increases, with its local minimum already above 35 eV
in Fig.\ \ref{figure2.ps}. The remaining case nearby 33 eV at $\Delta$
in Fig.\ \ref{figure1.ps} is not elucidated by Fig.\ \ref{figure2.ps};
one must increase the lattice constant to about 150\% to see that
the point with E$_u$ symmetry at X then falls below the point with
E$_g$ symmetry and gets connected with T$_{2g}$ at $\Gamma$, to
indicate that there also is accidental crossing of the two lines
with E symmetry nearby 34 eV on the $\Delta$ line in Fig.\ \ref{figure1.ps}.

An additional band of $\Gamma_2'$ (A$_{2u}$) symmetry appears in
calculations with a mixed basis set \cite{Kunz69} and an Augmented
Plane Wave basis (\cite{Page}, Fig.\ 1 at $1.9$ ryd) which is absent
in our Figs.\  \ref{figure1.ps} and \ref{figure2.ps}. If we augment
our LCAO orbital basis by a full set of five $f$-orbitals with
exponent $2.98632\cdot 10^{-1}$ at the F sites, it also emerges
from our calculations, $18.6$ eV above the $\Gamma_{1,c}$ bottom of the conduction band.
This basis set completion problem affects calculations of dielectric matrices
that sum over unoccupied states \cite{Reining}.

The choice of an $X_\alpha$-type LDA in Fig.\ \ref{figure1.ps} and \ref{figure2.ps}
is not to promote simplistic density functions, but for immediate
comparison with the graph in \cite{Zunger} and since this could serve
as a common standard reference of DFT implementations.

\subsection{Discussion}
\begin{figure}[h]
 
  \epsfig{scale=0.48,file=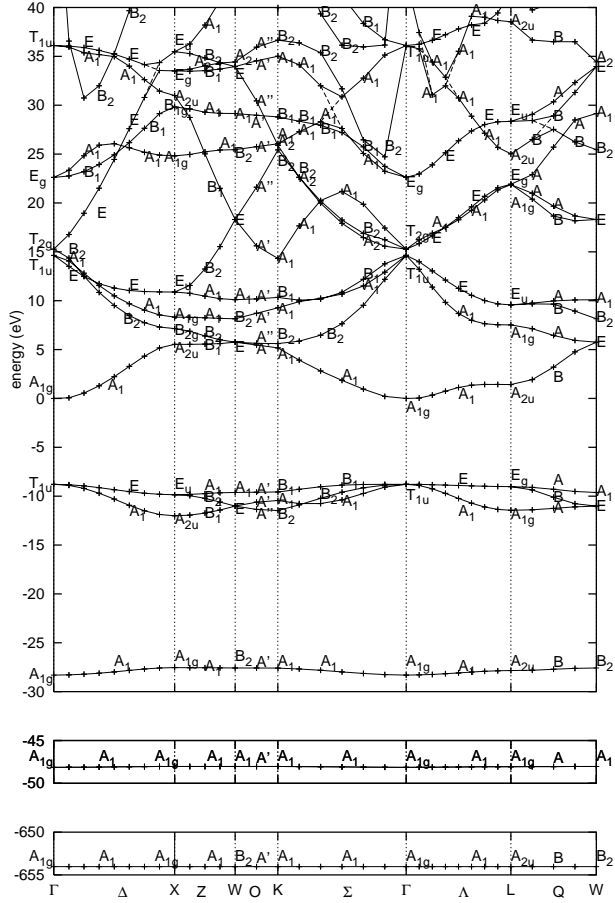,clip=}
\caption{
KS band-structure
of LiF at $a=7.5939a_0=4.01852$ {\AA} with
the $X_{\alpha}$ form of the XC energy ($\alpha=2/3$).
 Crosses mark energies in the BZ that were calculated by GTOFF on a grid of
$16\times 16\times 16$ points in the reciprocal unit cell.
The direct gap at $\Gamma$ is 8.79 eV\@.
\label{figure1.ps}
}
\end{figure}

\begin{figure}[h]
 
  \epsfig{scale=0.48,file=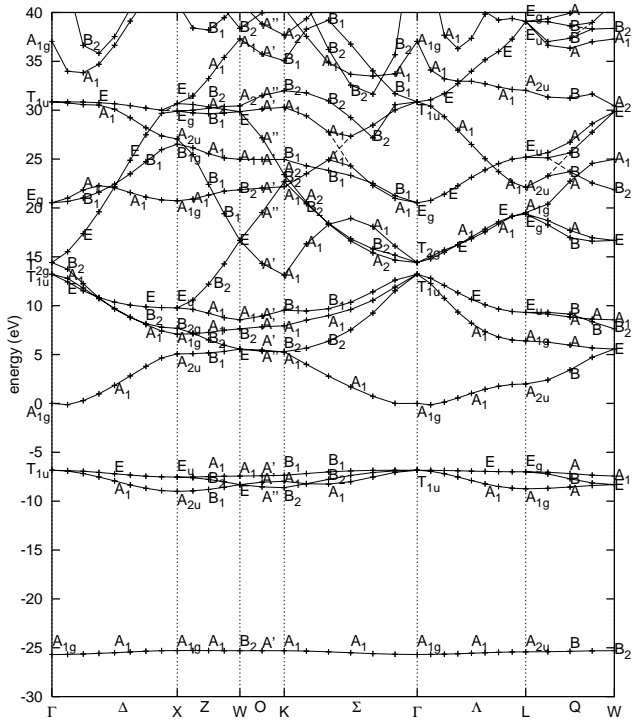,clip=}
\caption{
KS band-structure
of LiF as in Fig.\ \ref{figure1.ps} for a 10\% larger
lattice constant. As the lattice constant increases, the bottom of
the conduction band at $\Gamma$ becomes a local maximum with satellite
minima at the adjacent $\Delta$, $\Sigma$ and $\Lambda$ lines.
\label{figure2.ps}
}
\end{figure}

\begin{figure}[h]
 
  \epsfig{scale=0.48,file=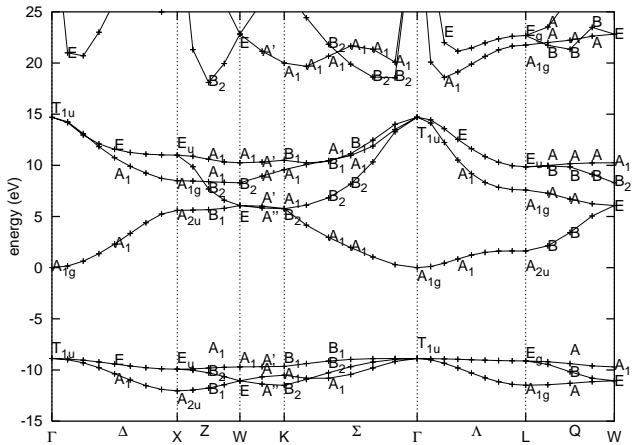,clip=}
\caption{
KS band-structure of LiF as in Fig.\ \ref{figure1.ps} but
the basis of Tab.\ \ref{gtos} replaced by a smaller basis of
Li 11$s$4$p$ (contracted 41111111/1111) and F 13$s$8$p$ (4111111111/2111111)
to illustrate the role of F $d$-orbitals.
\label{figure3.ps}
}
\end{figure}

A tour through the literature in roughly chronological order---strictly
confined to work that lists eigenvalues or band structures---leads to the following comparison:
\begin{itemize}

\item{\cite{Ewing}}
Fig.\ 5 is compatible with our results.

\item{\cite{Page}}
The L-labels in Fig.\ 1 suggest Page and Hygh use a setting with
Li in the center of coordinates: W$_1$ near $-0.45$ ryd should read W$_2'$.

\item{\cite{Chaney}}
The results in Tab.\ 1 and Fig.\ 3 are compatible with ours.
Chaney {\em et al.\/} did not include $d$-type basis functions,
which means the T$_{2g}$ point near 15 eV in our Fig.\ \ref{figure1.ps}
and bands that are attached to it disappear. In consequence they  neither show
a $\Gamma_{25c}'$ nor a X$_{3c}$ in Tab.\ 1\@.
Our Fig.\ \ref{figure3.ps} illustrates this case.
The $d$-orbitals are obviously a key ingredient to construct an even
function at $\Gamma$ in the conduction band, with otherwise only $s$-hybrids remaining
which have to stay orthogonal to the three $s$-type core orbitals.

\item{\cite{Drost}}
Figs.\ 1, 3 and 6:
the upper L$_2'$ should read L$_1$.
(The compatibility relations would
allow the connection $\Gamma_1$--$\Lambda_1$--L$_2'$.)
The equivalent correction in Tab.\ VIII is to exchange
L$_3'\rightarrow$ L$_2'$ with L$_3'\rightarrow$ L$_1$ in the
first column.

\item{\cite{Menzel,MenzelE}}
The results in Tab.\ 1 are compatible with those here in
the same sense as discussed for \cite{Chaney}.

\item{\cite{Brener73}}
Figs.\ 1 and 2:
the upper L$_2'$ should read L$_1$.
Tab.\ 1: replace L$_{2c}'$ by L$_{1c}$.

\item{\cite{Laramore}}
Figs.\ 3 and 4 are compatible with our analysis.

\item{\cite{Mickish}}
Fig.\ 1: the $1$ at X should read $4'$. A comment on the $3$ at
X and L is difficult because the order of their Hartree-Fock energy levels
might be switched in comparison to our DFT results.

\item{\cite{Euwema}}
The label X$_3$ in Tab.\ IV should read X$_4'$ where they
refer to their own results, as their basis does not include $d$-orbitals,
see the discussion of \cite{Chaney}).

\item{\cite{Brener75}}
Figs.\ 2 and 3: the uppermost X$_4'$ should read X$_1$.
(The compatibility relations would
allow the connection $\Gamma_{15}$--$\Delta_1$--X$_4'$.)
The uppermost L$_3'$ should read L$_3$. The middle L$_2'$
near 10 eV should read L$_1$.

\item{\cite{Kunz75}}
In the LiF part of Fig.\ 2, the $1$ at X should read $4'$.
(The compatibility relations would
allow the connection $\Gamma_1$--$\Delta_1$--X$_1$.)

\item{\cite{Pantel}}
This tight-binding analysis of the valence bands confirms our results.

\item{\cite{Piacentini}}
The lower L$_3'$ in Fig.\ 8 should be replaced by L$_2'$,
and the X$_1$ by X$_4'$.

\item{\cite{Zunger}}
The labels at L should read (bottom-up) L$_1$, L$_2'$, L$_3'$,
L$_1$, L$_2'$, L$_3$, L$_2'$, and L$_3'$ in Fig.\ 2.
All three cases of L$_3$ and L$_3'$ should split towards Q\@.
There should be no crossing at Q near 7 eV\@.
The $\Delta$-lines near 12 eV should twist to connect X$_3$
to $\Gamma_{25}'$ by $\Delta_2'$ and X$_1$ to $\Gamma_{15}$
by $\Delta_1$.
The labels $\Sigma_2$ and $\Sigma_3$ near
17 eV should probably be exchanged.
In Tab.\ 1,
all X$_4$, X$_5$ and L$_2$, also L$_{3v}$ should be primed.
X$_{1,c}$ in Tab.\ 1 should read X$_{3,c}$ where it refers to their
own calculations, which follows from
measuring the vertical distance in Fig.\ 2 and comparison with
the value of 0.710 hartree given in the table. (See again
the comment on \cite{Chaney}.)

\item{\cite{Antoci}}
Fig.\ 3 is compatible with our labeling scheme.

\item{\cite{Kunz82}}
In Fig.\ 2, the upper L$_2'$ should read L$_1$.

\item{\cite{Michiels}}
Fig.\ 1 is reproduced from \cite{Zunger}; hence the same comments hold.

\item{\cite{Wang}}
The following LDA lines in Fig.\ 1 should cross: Z lines near 41 eV;
O lines near 41 eV; $\Sigma$ lines near 30 eV and 39 eV;  $\Sigma$ lines
close to K near 35 eV\@.

\end{itemize}

\begin{acknowledgments}
We thank Peter Blaha for providing us with results of his independent
calculation. 
Jonathan Boettger's maintenance of the
GTOFF code was an indispensable foundation of the present analysis.
\end{acknowledgments}

\bibliography{lif.bib}

\begin{table}
\begin{ruledtabular}
\begin{tabular}{c|dd|dd|}
&\multicolumn{2}{c}{Li} & \multicolumn{2}{c}{F} \\
type & \text{exponent} & \text{coefficient} & \text{exponent} & \text{coefficient} \\
\hline
s  & 7.471871\cdot 10^4  & 0.000006 & 4.007457\cdot 10^5  & 0.0000114 \\
   & 1.11883\cdot 10^4   & 0.000044 & 6.001308\cdot 10^4  & 0.0000904 \\
   & 2.546139\cdot 10^3  & 0.000232 & 1.36574\cdot 10^4   & 0.0004719 \\
   & 7.211808\cdot 10^2  & 0.000978 & 3.868237\cdot 10^3  & 0.001991 \\
   & 2.352752\cdot 10^2  & 0.003545 & 1.261848\cdot 10^3  & 0.0071938 \\
   & 8.493513\cdot 10^1  & 0.011395 & 4.554532\cdot 10^2  & 0.0229084 \\
   & 3.312063\cdot 10^1  &          & 1.775331\cdot 10^2  &           \\
   & 1.371652\cdot 10^1  &          & 7.345899\cdot 10^1 &  \\
   & 5.941083\cdot 10^0  &          & 3.181364\cdot 10^1 & \\
   & 2.651974\cdot 10^0  &          & 1.426084\cdot 10^1 & \\
   & 1.209217\cdot 10^0  &          & 6.478246\cdot 10^0 & \\
   & 5.56503\cdot 10^{-1}&          & 2.627737\cdot 10^0  & \\
   & 2.37586\cdot 10^{-1}&          & 1.184078\cdot 10^0  & \\
   &                       &          & 5.05251\cdot 10^{-1}& \\
   &                       &          & 2.1008\cdot 10^{-1} & \\
\hline
p  & 3.337598\cdot 10^1  &          & 6.600857\cdot 10^2  & 0.0001863 \\
   & 7.907316\cdot 10^0  &          & 1.564625\cdot 10^2  & 0.0015866 \\
   & 2.50156\cdot 10^0   &          & 5.065363\cdot 10^1  & 0.008512 \\
   & 8.8178\cdot 10^{-1} &          & 1.908092\cdot 10^1  & \\
   & 3.30977\cdot 10^{-1}&          & 7.872742\cdot 10^0 & \\
   &                       &          & 3.449055\cdot 10^0 & \\
   &                       &          & 1.544895\cdot 10^0 & \\
   &                       &          & 6.86529 \cdot 10^{-1}& \\
   &                       &          & 2.98632 \cdot 10^{-1}& \\
   &                       &          & 1.24571 \cdot 10^{-1}& \\
\hline
d &                       &           & 2.98632 \cdot 10^{-1}& \\
  &                       &           & 1.24571 \cdot 10^{-1}& \\
\end{tabular}
\end{ruledtabular}
\caption{
Exponentials and contraction coefficients of GTO's centered at F and Li,
derived from Partridge's 13$s$9$p$ and 15$s$10$p$ sets \cite{Partr,PartrMemo}.
\label{gtos}
}
\end{table}

\begin{table}
\begin{ruledtabular}
\begin{tabular}{cc|cc|cc|cc|cc|cc|cc|cc}
\multicolumn{2}{c}{W} & \multicolumn{2}{c}{L} &\multicolumn{2}{c}{X} 
&\multicolumn{2}{c}{$\Gamma$} &\multicolumn{2}{c}{K($\Sigma$)} 
&\multicolumn{2}{c}{Z} 
&\multicolumn{2}{c}{$\Delta$} & \multicolumn{2}{c}{$\Lambda$}\\
\multicolumn{2}{c}{$\bar 42m$} & \multicolumn{2}{c}{$\bar 3m$} &
\multicolumn{2}{c}{$4/mmm$} 
&\multicolumn{2}{c}{$m3m$} &\multicolumn{2}{c}{$mm2$} &\multicolumn{2}{c}{$mm2$} 
&\multicolumn{2}{c}{$4mm$} & \multicolumn{2}{c}{$3m$}\\
\hline
A$_1$ & W$_1$  & A$_{1g}$ & L$_1$  &  A$_{1g}$ & X$_1$
      & A$_{1g}$ & $\Gamma_1$ & A$_1$ & K$_1$  & A$_1$ & Z$_1$ & A$_1$ & $\Delta_1$ 
      & A$_1$ & $\Lambda_1$ \\
B$_1$ & W$_1'$ & A$_{2g}$ & L$_2$  &  B$_{1g}$ & X$_2$ 
      & A$_{2g}$ & $\Gamma_2$ & A$_2$ & K$_2$ & A$_2$ & Z$_2$ &  B$_1$ & $\Delta_2$ 
      & A$_2$ & $\Lambda_2$ \\
A$_2$ & W$_2$  & E$_g$ & L$_3$     &  B$_{2g}$ & X$_3$ 
      & E$_g$ & $\Gamma_{12}$ & B$_2$ & K$_3$ & B$_1$ & Z$_3$ & B$_2$ & $\Delta_2'$
      & E & $\Lambda_3$ \\
B$_2$ & W$_2'$ & A$_{1u}$ & L$_1'$ &  A$_{2g}$ & X$_4$ 
      & T$_{1g}$ & $\Gamma_{15}'$ &  B$_1$ & K$_4$ & B$_2$ & Z$_4$ & A$_2$ & $\Delta_1'$
      & & \\
E     & W$_3$  & A$_{2u}$ & L$_2'$ &  A$_{1u}$ & X$_1'$
      & T$_{2g}$ & $\Gamma_{25}'$ &  & & & & E & $\Delta_5 $
      & & \\
      &        & E$_u$ &    L$_3'$ &  B$_{1u}$ & X$_2'$ 
      & A$_{1u}$ & $\Gamma_1'$ & & & & & &
      & & \\
      &        &       &           &  B$_{2u}$ & X$_3'$ 
      & A$_{2u}$ & $\Gamma_2'$ & & & & & &
      & & \\
      &        &       &           &  A$_{2u}$ & X$_4'$ 
      & E$_u$ & $\Gamma_{12}'$ & & & & & &
      & & \\
      &        &       &           &  E$_g$ & X$_5$     
      & T$_{1u}$ & $\Gamma_{15}$  & & & & & &
      & & \\
      &        &       &           &  E$_u$ & X$_5'$  
      & T$_{2u}$ & $\Gamma_{25} $ & & & & & &
      & & \\
\end{tabular}
\end{ruledtabular}
\caption{
The interpretation of the point group symbols of \cite{Bradley}
in terms of those used in \cite{Bouckaert}, obtained by
searching for matching lines in the character tables.
There is only one table for $mm2$ in \cite[p.\ 58]{Bradley} and only
one common table for Z and K in \cite[Tab.\ VI]{Bouckaert}, but 
the $\sigma_x$ and $\sigma_z$ operations are in different classes
in \cite[Tab.\ VI]{Bouckaert}, which leads to two separate columns
for Z and K here.
\label{pglabels}}
\end{table}

\end{document}